\documentclass[conference]{IEEEtran}
\ifCLASSINFOpdf
\else
\fi

\usepackage{graphicx}
\usepackage{graphics}
\usepackage{subfigure}
\usepackage{url}
\usepackage{verbatim} % Block Comments
\usepackage{multicol}
\usepackage[small]{caption}
\usepackage{array}
\usepackage{margins}
\usepackage{psfrag}

\usepackage{amsthm}
\usepackage{amsmath, amssymb}
\usepackage{amsfonts}

\newcommand{\rvx}{\mathsf{x}}

\newcommand{\rvs}{\mathsf{s}}

\newcommand{\cC}{{\mathcal C}}

\newcommand{\al}{\alpha}
\newcommand{\cp}{\check{p}}
\newcommand{\tp}{\tilde{p}}
\newcommand{\bp}{\mathbf{p}}
\newcommand{\bs}{\mathbf{s}}
\newcommand{\bq}{\mathbf{q}}
\newcommand{\bd}{\mathbf{d}}
\newcommand{\ts}{\tilde{s}}
\newcommand{\CapTau}{\mathcal{T}}

\newcommand{\pvec}{\mathbf{p}}
\newcommand{\svec}{\mathbf{s}}

\newcommand{\qvec}{\mathbf{q}}
\newcommand{\dvec}{\mathbf{d}}
\newtheorem{thm}{Theorem}
\newtheorem{lem}{Lemma}

\newtheorem{definition}{Definition}

\begin{document}

% Paper Title
\title{Diversity Embedded Streaming Erasure Codes (DE-SCo): Constructions and Optimality}

\author{\IEEEauthorblockN{ Ahmed Badr and Ashish Khisti}
\IEEEauthorblockA{Dept. of Electrical and Computer Engineering\\
University of Toronto\\
Toronto, ON, M5S 3G4, Canada\\
Email: \{abadr, akhisti\}@comm.utoronto.ca}
\and
\IEEEauthorblockN{Emin Martinian}
\IEEEauthorblockA{Signals, Information and Algorithms Laboratory\\
Massachusetts Inst. of Technology\\
Cambridge, MA, 02139\\
Email: emin@alum.mit.edu}}


\maketitle

\begin{abstract}
Streaming erasure codes encode a source stream to guarantee that each source packet is recovered within a fixed delay at the receiver over a burst-erasure channel. This paper introduces  \emph{diversity embedded streaming erasure codes} (DE-SCo), that provide a flexible tradeoff between the channel quality and receiver delay. When the channel conditions are good, the source stream is recovered with a  low delay, whereas when the channel conditions are poor the source stream is still recovered, albeit with a larger delay.  Information theoretic analysis of the underlying burst-erasure broadcast channel reveals that  DE-SCo  achieve the minimum possible delay for the weaker user, without sacrificing the performance of the stronger user.  A larger class of multicast streaming erasure codes (MU-SCo) that achieve optimal tradeoff between rate, delay and erasure-burst length is also constructed. %\boldmath
\end{abstract}

\IEEEpeerreviewmaketitle

\vspace{-1em}
\section{Introduction}
Forward error correction codes  designed for streaming sources require that (a) the channel input stream be produced sequentially from the source stream (b) the decoder sequentially reconsructs the source stream as it observes the channel output. In contrast, traditional error correction codes such as maximum distance separable (MDS) codes map blocks of data to a codeword and the decoder waits until the entire codeword is received before the source data can be reproduced. Rateless codes such as the digital fountain codes also do not form ideal streaming codes. First they require that the entire source data be available before the output stream is reproduced. Secondly they provide no guarantees on the sequential reconstruction of the source stream. Non-block codes such as convolutional codes in conjunction with sequential decoding can be designed for low delay applications~\cite{Jelinek}. However to our best knowledge, these constructions need to be optimized through a numerical search for finite constraint lengths. Low-delay codes with feedback are recently studied in~\cite{Sahai,JayKumar} while compression of streaming sources is studied in~\cite{Draper}. 

In~\cite[Chapter 8]{Martinian_Thesis} a new class of codes,~ \emph{streaming erasure codes} (SCo) are proposed.  The encoder observes a semi-infinite source stream --- one packet is revealed in each time slot  --- and maps it to a coded output stream of rate $R$. The channel is modelled as a burst-erasure channel. Starting at an arbitrary time, it introduces an erasure-burst of maximum length $B$. The decoder is required to reconstruct each source packet with a maximum delay $T$. A fundamental relationship between $R$, $B$ and $T$ is established and  SCo codes are constructed that achieve this tradeoff. We emphasize that the parity check symbols in these  constructions involve a careful combination of source symbols.
In particular, random linear combinations, popularly used in e.g., network coding, do not attain the optimal performance.

The  SCo framework however requires that the value of $B$ and $T$ be known apriori. In practice this forces a conservative design i.e., we design the code for the worst case $B$ thereby incurring a higher overhead (or a larger delay) even when the channel is relatively good. Moreover there is often a flexibility in the  allowable delay. Techniques such as adaptive media playback~\cite{Girod} have been designed to tune the play-out rate as a function of the received buffer size to deal with a temporary increase in delay. Hence it is  not desirable to have to fix $T$ during the design stage either.

The streaming codes introduced in this work do not commit apriori to a specific delay. Instead they realize a delay that depends on  the channel conditions. At an information theoretic level, our setup extends the point-to-point link in~\cite{Martinian_Thesis} to a multicast model --- there is one source stream and two receivers. The channel for each receiver introduces an erasure-burst of length $B_i$ and each receiver can tolerate a delay of $T_i$.  We investigate multicast streaming erasure code (MU-SCo) constructions  that achieve the maximum rate under these constraints. We primarily focus on a particular subclass  --- diversity embedded streaming erasure codes (DE-SCo). These codes modify a single user SCo such that the resulting code can support a second user, whose channel introduces a larger erasure-burst, without sacrificing the performance of the first user.
Our construction  embeds new parity checks in an SCo code in  a manner such that (a) no interference is caused to the stronger (and low delay) user and (b) the weaker user can use some of the parity checks of the stronger user as side information to recover part of the source symbols. DE-SCo constructions  outperform  baseline schemes that  simply concatenate the single user SCo for the two users. An information theoretic converse establishes that DE-SCo achieves the minimum possible delay for the weaker receiver without sacrificing the performance of the stronger user. Finally all our code constructions can be encoded and decoded with a polynomial time complexity in $T$ and $B$.
\section{System Model}
\label{Background}
\begin{figure}
\centering
\includegraphics[scale=.5, trim = 15mm 250mm 10mm 10mm, clip]{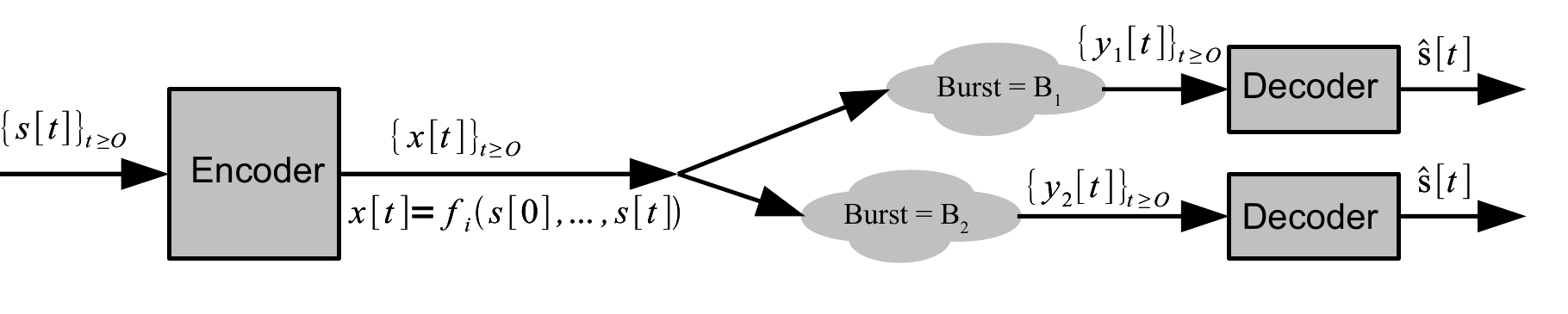}
\caption{The source stream $\{\rvs[i]\}$ is causally mapped into an output stream $\{\rvx[i]\}$. Both the receivers observe these packets via their channels. The channel introduces an erasure-burst of length $B_i$, and each receiver tolerates a delay of $T_i$, for $i=1,2$.}
\label{ProblemFormulationFigure}
\vspace{-1 em}
\end{figure}
The transmitter encodes a stream of source packets $\{ s[t] \}_{t \geq 0}$ intended to be received at two receivers as shown in Fig.~\ref{ProblemFormulationFigure}. The channel packets $\{ x[t] \}_{t \geq 0}$ are produced causally from the source stream, $x[t] = f_t (s[0],\dots,s[t])$.

%\vspace{-1em}

%\begin{equation}
%\label{Code_Function}

%\end{equation}
The channel of receiver $i$ introduces an erasure-burst of length $B_i$ i.e., the channel output at receiver $i$ at time $t$ is given by

\vspace{-1em}

\begin{equation}
\label{Channel_Function_Multi}
y_i[t] = 
\left\{
\begin{array}{ll}
\star & t \in [j_i,j_i + B_i - 1] \\
x[t] & \text{otherwise}
\end{array}
\right.
\end{equation}
for $i=1,2$ and for some $j_i \ge 0$. Furthermore, user $i$ tolerates a delay of $T_i$, i.e., there exists a sequence of decoding functions $\gamma_{1t}(.)$ and $\gamma_{2t}(.)$ such that
\begin{equation}
\label{Decoders}
\hat{s}_i[t] = \gamma_{it} (y_1[0],y_1[1],\dots,y_1[i+T_i]), \qquad i=1,2,
\end{equation}
and $\text{Pr} (s_i[t] \neq \hat{s}_i[t]) = 0, \; \; \; \;	\forall t \geq 0, \; \;$.
%\begin{equation}
%\label{Prob_Error_Multi}

%\end{equation}
The source stream is an i.i.d.\ process; each source symbol is sampled from a distribution $p_\rvs(\cdot)$. The rate of the multicast code is defined as ratio of the (marginal) entropy of the source symbol to the (marginal) entropy of each channel symbol i.e., $R = H(\rvs)/H(\rvx)$. An \emph{optimal multicast streaming erasure code (MU-SCo)} achieves the maximum rate for a given choice of $(B_i, T_i)$. Of particular interest is the following subclass.
\begin{definition}[Diversity Embedded Streaming Erasure Codes (DE-SCo)] 
\label{defn:DE-SCo}
Consider the multicast model in Fig.~\ref{ProblemFormulationFigure} where the channels of the two receivers introduce an erasure burst of lengths $B_1$ and $B_2$ respectively with $B_1 < B_2$.  A DE-SCo is a rate $R =\frac{T_1}{T_1+B_1}$ MU-SCo construction that achieves a delay $T_1$ at receiver $1$ and supports  receiver $2$ with delay $T_2$. An optimal DE-SCo minimizes the delay $T_2$ at receiver $2$ for given values of $B_1$, $T_1$ and $B_2$. \end{definition}

Our setup generalizes the point-to-point case in~\cite[Chapter 8]{Martinian_Thesis}  where single user SCo codes for parameters $(B_1,T_1)$ achieve the streaming capacity $\frac{T_1}{T_1+B_1}$. An optimal MU-SCo construction, attains the maximum rate for fixed parameters $(B_i,T_i)$.  An optimal DE-SCo fixes the rate to the capacity of user 1, and  supports user $2$ with the minimum possible delay $T_2$.   

Note that our model only considers a single erasure burst on each channel. As is the case with (single user) SCo, our constructions correct multiple erasure-bursts  separated sufficiently apart. Also we only consider the erasure channel model. It naturally arises when these codes are implemented in application layer multimedia encoding. More general channel models can be transformed into an erasure model by applying an appropriate inner code~\cite[Chapter 7]{Martinian_Thesis}.

\section{Example}
\label{sec:Example}
\begin{table*}[t]
\centering
\subfigure[SCo Construction for $(B,T) = (1,2)$]{
\label{Code1224_a}
\begin{tabular}{|c|c|c|c|c|c|}
\hline
\fbox{$s_0[i-1]$} & $s_0[i]$ & $s_0[i+1]$ & $s_0[i+2]$ & $s_0[i+3]$ & $s_0[i+4]$ \\
$s_1[i-1]$ & \fbox{$s_1[i]$} & $s_1[i+1]$ & $s_1[i+2]$ & $s_1[i+3]$ & $s_1[i+4]$ \\
\hline
$s_0[i-3] \oplus s_1[i-2]$ & $s_0[i-2] \oplus s_1[i-1]$ & \fbox{$s_0[i-1] \oplus s_1[i]$} & $s_0[i] \oplus s_1[i+1]$ & $s_0[i+1] \oplus s_1[i+2]$ & $s_0[i+2] \oplus s_1[i+3]$\\
\hline
\end{tabular}
}
\subfigure[SCo Construction for $(B,T) = (2,4)$]{
\begin{tabular}{|c|c|c|c|c|c|}
\hline
\fboxrule=1pt
\fbox{$s_0[i-1]$} & $s_0[i]$ & $s_0[i+1]$ & $s_0[i+2]$ & $s_0[i+3]$ & $s_0[i+4]$ \\
$s_1[i-1]$ & $s_1[i]$ & \fboxrule=1pt \fbox{$s_1[i+1]$} & $s_1[i+2]$ & $s_1[i+3]$ & $s_1[i+4]$ \\
\hline
$s_0[i-5] \oplus s_1[i-3]$ & $s_0[i-4] \oplus s_1[i-2]$ & $s_0[i-3] \oplus s_1[i-1]$ & $s_0[i-2] \oplus s_1[i]$ & \fboxrule=1pt \fbox{$s_0[i-1] \oplus s_1[i+1]$} & $s_0[i] \oplus s_1[i+2]$\\
\hline
\end{tabular}
\label{Code1224_b}
}
\subfigure[Cc-SCo for $\{ (B_1,T_1),(B_2,T_2) \} = \{ (1,2),(2,4) \}$]{
\begin{tabular}{|c|c|c|c|c|c|}
\hline
\fboxrule=1pt \fbox{ \fboxrule=0.5pt \fbox{$s_0[i-1]$} } & $s_0[i]$ & $s_0[i+1]$ & $s_0[i+2]$ & $s_0[i+3]$ & $s_0[i+4]$ \\
$s_1[i-1]$ & \fbox{$s_1[i]$} & \fboxrule=1pt \fbox{$s_1[i+1]$} & $s_1[i+2]$ & $s_1[i+3]$ & $s_1[i+4]$ \\
\hline
$s_0[i-3] \oplus s_1[i-2]$ & $s_0[i-2] \oplus s_1[i-1]$ & \fbox{$s_0[i-1] \oplus s_1[i]$} & $s_0[i] \oplus s_1[i+1]$ & $s_0[i+1] \oplus s_1[i+2]$ & $s_0[i+2] \oplus s_1[i+3]$\\\hline
$s_0[i-5] \oplus s_1[i-3]$ & $s_0[i-4] \oplus s_1[i-2]$ & $s_0[i-3] \oplus s_1[i-1]$ & $s_0[i-2] \oplus s_1[i]$ & \fboxrule=1pt \fbox{$s_0[i-1] \oplus s_1[i+1]$} & $s_0[i] \oplus s_1[i+2]$\\
\hline
\end{tabular}
\label{Code1224_c}
}
\caption{Single user SCo constructions are shown in the upper two figures. The Cc-SCo construction in the last figure repeats the parity checks of the two single user codes to simultaneously satisfy both the users. This baseline approach is in general sub-optimal }
\label{Code1224}
\end{table*}

\normalsize

We first highlight our results via a numerical example: $(B_1,T_1)= (1,2)$ and $(B_2,T_2) = (2,4)$. Single user SCo constructions from~\cite{Martinian_Thesis,Martinian_Trott} for both users are illustrated in  Table~\ref{Code1224}(a) and~\ref{Code1224}(b) respectively. In each case, the source symbol $\rvs[i]$ is split into two  sub-symbols $(\rvs_0[i],\rvs_1[i])$ and the channel symbol $\rvx[i]$ is obtained by concatenating the source symbol $\rvs[i]$ with a parity check symbol $p[i]$. In the $(1,2)$ SCo construction, parity check symbol $p_A[i] = s_1[i-1]\oplus s_0[i-2]$ is generated by combining the source sub-symbols diagonally across the source stream as illustrated with the rectangular boxes. For the $(B,T) = (2,4)$, the choice $p_B[i] = s_1[i-2]\oplus s_0[i-4]$ is similar, except that an interleaving of step of size $2$ is applied before the parity checks are produced. 

How can we construct a single code that simultaneously supports both $\{(1,2),(2,4)\}$? The first approach is to concatenate the two parity check streams as shown in Table.~\ref{Code1224}(c). Each receiver ignores the parity check rows of the other receiver and performs single user decoding. However such  a concatenated streaming code (Cc-SCo) achieves a rate of $1/2$, whereas the optimal MU-SCo construction achieves a rate of $3/5$. Before specifying the optimal MU-SCo code construction, let us first consider designing a DE-SCo (see Def.~\ref{defn:DE-SCo}) for this setup. Recall that this code, of rate $2/3$, satisfies  $(B_1,T_1) = (1,2)$,  also achieves minimum delay $T_2$ for $B_2 = 2$. 

\begin{table*}[t]
\centering
\subfigure[IA-SCo Code Construction for $(B_1,T_1) = (1,2)$ and $(B_2,T_2) = (2,6)$]{
\begin{tabular}{|c|c|c|c|c|c|}
\hline
$s_0[i-1]$ & $s_0[i]$ & $s_0[i+1]$ & $s_0[i+2]$ & $s_0[i+3]$ & $s_0[i+4]$ \\
$s_1[i-1]$ & $s_1[i]$ & $s_1[i+1]$ & $s_1[i+2]$ & $s_1[i+3]$ & $s_1[i+4]$ \\
\hline
$s_0[i-3] \oplus s_1[i-2]$ & $s_0[i-2] \oplus s_1[i-1]$ & $s_0[i-1] \oplus s_1[i]$ & $s_0[i] \oplus s_1[i+1]$ & $s_0[i+1] \oplus s_1[i+2]$ & $s_0[i+2] \oplus s_1[i+3]$ \\
$\oplus$ & $\oplus$ & $\oplus$ & $\oplus$ & $\oplus$ & $\oplus$ \\
$s_0[i-7] \oplus s_1[i-5]$ & $s_0[i-6] \oplus s_1[i-4]$  & $s_0[i-5] \oplus s_1[i-3]$ & $s_0[i-4] \oplus s_1[i-2]$ & $s_0[i-3] \oplus s_1[i-1]$ & $s_0[i-2] \oplus s_1[i]$\\
\hline
\end{tabular}}
\subfigure[DE-SCo Code Construction for $(B_1,T_1) = (1,2)$ and $(B_2,T_2) = (2,5)$]{\begin{tabular}{|c|c|c|c|c|c|}
\hline
\fbox{$s_0[i-1]$} & \fboxrule=1pt \fbox{$s_0[i]$} & $s_0[i+1]$ & $s_0[i+2]$ & $s_0[i+3]$ & $s_0[i+4]$ \\
\fboxrule=1pt \fbox{$s_1[i-1]$} & \fbox{$s_1[i]$} & $s_1[i+1]$ & $s_1[i+2]$ & $s_1[i+3]$ & $s_1[i+4]$ \\
\hline
$\left\{ s_0[i-3] \oplus s_1[i-2] \right.$ & $\left\{s_0[i-2] \oplus s_1[i-1] \right.$ & \fbox{$s_0[i-1] \oplus s_1[i]$} & $\left\{s_0[i] \oplus s_1[i+1]\right.$ & $\left\{s_0[i+1] \oplus s_1[i+2]\right.$ & $s_0[i+2] \oplus s_1[i+3]$ \\
$\oplus$ & $\oplus$ & $\oplus$ & $\oplus$ & $\oplus$ & $\oplus$ \\
$\left. s_1[i-6] \oplus s_0[i-5]\right\}$ & $\left. s_1[i-5] \oplus s_0[i-4]\right\}$  & $\left. s_1[i-4] \oplus s_0[i-3]\right.$ & $\left. s_1[i-3] \oplus s_0[i-2]\right\}$ & $\left. s_1[i-2] \oplus s_0[i-1]\right\}$ & \fboxrule=1pt \fbox{$s_1[i-1] \oplus s_0[i]$}\\
\hline
\end{tabular}}
\caption{Rate $2/3$ code constructions that satisfy user 1 with $(B_1,T_1) = (1,2)$ and user $2$ with $B_2 = 2$.}
\label{Code1225}
\end{table*}

In Table~\ref{Code1225}(a) we illustrate a rate $R =2/3$  code that achieves $T_2=6$. It is obtained by shifting the parity checks of the SCo code in Table~\ref{Code1224}(b) to the right by two symbols and combining with the parity checks of the SCo code in Table~\ref{Code1224}(a) i.e., $q[i] = p_A[i] \oplus p_B[i-2]$. Note that parity check symbols $p_B[\cdot]$ do not interfere with the parity checks of user $1$ i.e., when $s[i]$ is erased, receiver 1 can recover $p_A[i+1]$ and $p_A[i+2]$ from $q[i+1]$ and $q[i+2]$ respectively by cancelling $p_B[\cdot]$ that combine with these symbols. It then recovers $s[i]$.  Likewise if $s[i]$ and $s[i-1]$ are erased, then receiver $2$ recovers $p_B[i+1],\ldots, p_B[i+4]$ from $q[i+3],\ldots,q[i+6]$ respectively by cancelling out the interfering $p_A[\cdot]$, thus yielding $T_2 = 6$. 

The interference avoidance strategy illustrated above is sub-optimal.  Table.~\ref{Code1225}(b) shows the DE-SCo construction that achieves the minimum possible delay of $T_2 = 5$. In this construction we first construct the parity checks $\cp_B[i] = s_1[i-2]\oplus s_0[i-1]$ by combining the source symbols along the opposite diagonal of the $(1,2)$ SCo code in Table~\ref{Code1224}(a). Note that $x(i) = (s[i], \cp_B[i])$ is also a single user $(1,2)$ SCo code. We then shift the parity check stream to the right by $T+B = 3$ symbols and combine with $p_A[i]$ i.e., $q[i] = p_A[i] \oplus \cp_B[i-3]$. In the resulting code, receiver $1$ is still able to cancel the effect of $\cp_B[\cdot]$ as before and achieve $T_1=2$. Furthermore at receiver $2$ if $s[i]$ and $s[i-1]$ are erased, then observe that receiver $2$ obtains $s_0[i]$ and $s_0[i-1]$ from $q[i+2]$ and $q[i+3]$ respectively and $s_1[i-1]$ and $s_1[i]$ from $q[i+4]$ and $q[i+5]$ respectively, thus yielding $T_2=5$ symbols. 

Finally, the   $R=3/5$, MU-SCo $\{(1,2),(2,4)\}$ construction is as follows. Split  each source symbol $s[i]$ into six sub-symbols $s_0[i],\ldots, s_5[i]$ and construct an expanded source sequence $\ts[\cdot]$ such that $\ts[2i] = (s_0[i],s_1[i],s_2[i])$ and $\ts[2i+1]=(s_3[i],s_4[i],s_5[i])$. There exists a $\{(2,3),(4,8)\}$ DE-SCo code (see Thm.~\ref{thm:DE-SCo}) that we apply to $\ts[\cdot]$ to produce the parity checks $\tp[\cdot]$ and transmit $p[i] = (\tp[2i],\tp[2i+1])$ along with $s[i]$ at time $i$. Due to the properties of the DE-SCo construction and our source expansion, the resulting code corrects a single erasure with a delay of $2$ symbols and an erasure-burst of length $2$ with a delay of $4$. An explicit construction is provided in the full paper~\cite{FullPaper} due to space constraints. 

\section{ Construction of DE-SCo}
In this section we describe the DE-SCo construction. We rely on several properties of the single user SCo; see~\cite[Chapter 7-8]{Martinian_Thesis},\cite{Martinian_Trott} for the background;~\cite{FullPaper} for a detailed proof.
\begin{thm}
\label{thm:DE-SCo}
Let $(B_1,T_1)= (B,T)$ and suppose $B_2 = \al B$ where $\al$ is any integer that exceeds $1$. The minimum possible delay for any code of rate $R=\frac{T}{T+B}$ is
\begin{equation}
\label{eq:DE-SCoT2}T_2^\star = \al T + B,\end{equation}and is achieved by the optimal DE-SCo construction.
\end{thm}The converse in Theorem~\ref{thm:DE-SCo} states that  the rate of any $\{(B,T),(B_2,T_2)\}$ MU-SCo with $T_2 < T^\star_2$ is strictly below $R$. We establish this by constructing a periodic burst-erasure channel in which  every period of $(\al -1)B + T_2$ symbols consists of a sequence of $\al B$ erasures followed by  a sequence of non-erased symbols. Following a similar line of reasoning as in~\cite[Theorem 2]{Khisti} we can show  that a MU-SCo code  corrects all erasures on this channel. The rate of the MU-SCo code then must not exceed $1 - \frac{\al B}{(\al-1)B + T_2}$ which is less than $R$ if $T_2 < T_2^\star$.

For achievability, $T_2^\star$ in~\eqref{eq:DE-SCoT2} we construct the following code:\begin{itemize}
\item Let $\cC_1$  be the single user $(B,T)$ SCo obtained by splitting each source symbol $s[i]$ into $T$ sub-symbols $(s_1[i],\ldots,s_T[i])$ and producing $B$ parity check sub-symbols $\bp^A = (p^A_1[i],\ldots, p^A_B[i])$ at each time by combining the source sub-symbols along the main diagonal i.e.,\begin{equation}\label{eq:pA}p^A_k[i] = A_{k}(s_1[i-T-(k-1)],\ldots, s_T[i-1-(k-1)])\end{equation}

\item  Let $\cC_2$ be a  $((\al-1)B_1, (\al-1)T_1)$ SCo also obtained by splitting the source symbols into $T$ sub-symbols $(s_1[i],\ldots,s_T[i])$ and then constructing a total of $B$ parity checks $\bp^B = (p^B_1[i],\ldots, p^B_B[i])$ by combining the source sub-symbols along the opposite diagonal and with an interleaving step of size $\ell= (\al-1)$ i.e.,\begin{equation}\label{eq:pB}p^B_k[i] = B_k(s_1[i-\ell-(k-1)\ell],\ldots, s_T[i- \ell T - (k-1)\ell]).\end{equation}

\item Introduce a shift $\Delta = T+B$ in the stream $p_B[\cdot]$ and combine with the parity check stream $p^A[\cdot]$ i.e.,  $\bq[i] = \bp^A[i]\oplus \bp^B[i-\Delta]$. The output symbol at time $i$ is $x[i] = (s[i],\bq[i])$ 
\end{itemize}

Since there are $B$ parity check sub-symbols for every $T$ source sub-symbols it follows that the rate of the code is $\frac{T}{T+B}$.  
\subsubsection*{Decoding at User $1$}Suppose that the symbols at time $i-B,\ldots, i-1$ are erased by the channel of user $1$. User 1 first recovers parity checks $\bp^A[i],\ldots, \bp^A[i+T-1]$ from $\bq[i],\ldots, \bq[i+T-1]$ by cancelling the parity checks $\bp^B[\cdot]$ that combine with $\bp^A[\cdot]$ in this period. Indeed at time $i+T-1$ the interfering parity check is $\bp^B[i+T-\Delta-1]=\bp^B[i-B-1]$, which clearly depends on the (non-erased) source symbols before time $i-B$. All parity checks $\bp^B[\cdot]$ before this time are also non-interfering. The erased source symbols can be recovered from $\bp^A[i],\ldots, \bp^A[i+T-1]$ by virtue of code $\cC_1$.

\begin{figure}
		\centering
		\includegraphics[scale=0.35, trim = 0mm 90mm 0mm 25mm, clip]{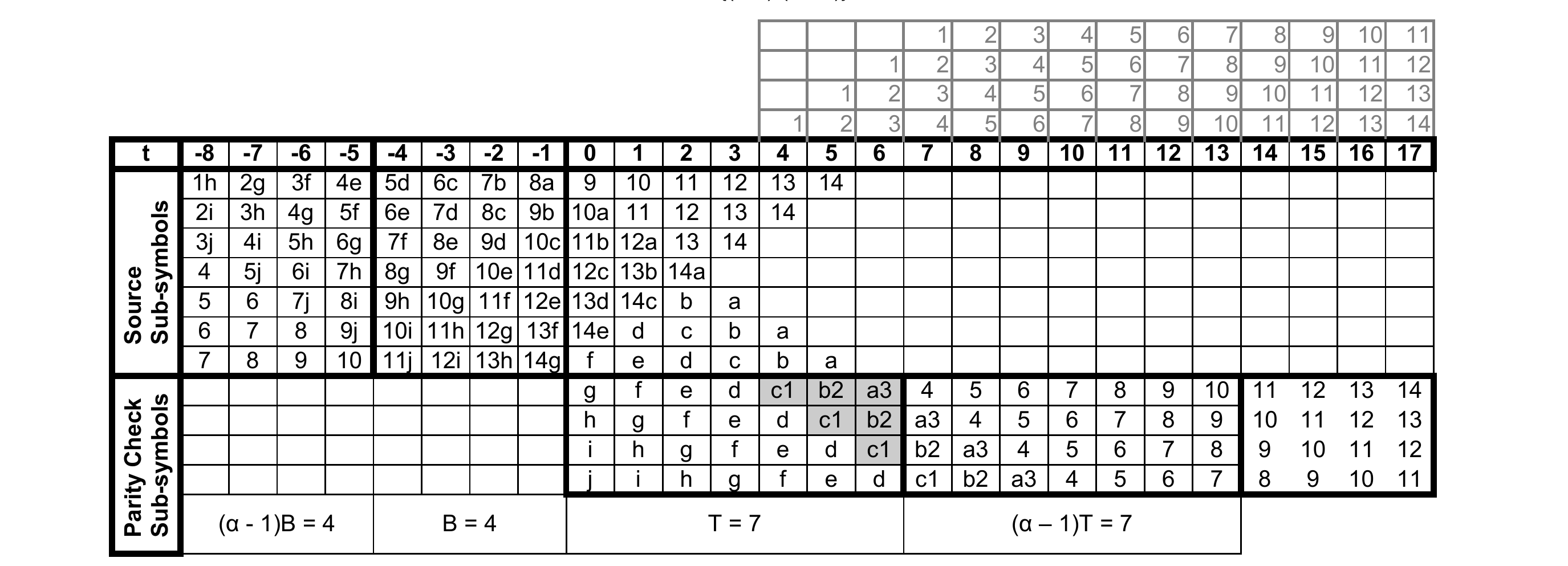}
		%\resizebox{\columnwidth}{!}{\includegraphics[trim = 0mm 0mm 0mm 90mm, clip]{Regions_Figure.pdf}}
		\caption{Decoding for DE-SCo  $\{(4,7),(8,18)\}$.}
		\label{DecodingExample}
\end{figure}

\subsubsection*{Decoding at User $2$}Suppose that the symbols at times $i-1,\ldots, i-\al B$ are erased for receiver 2. The decoding proceeds as follows:
\begin{enumerate}
\item[Step 1] At times $t \ge i+T$, the decoder recovers parity check $\bp^B[t]$ from $\bq[t]$ by cancelling the parity checks $\bp^A[t]$ which depend only on (non-erased) source symbols at time $i$ or later. 
\item[Step 2] At times $i\le t < i+T$ the decoder uses $\bq[t]$ and at times $i+T \le t < \CapTau\stackrel{\Delta}{=} i-\al B + T^\star_2$, the decoder uses $\bp^B[t]$ (obtained in Step 1), to recover symbols $\{\bs^N[\tau]\}_{\tau=i-\al B}^{i-1}$ where $ \bs^N[\tau]=(s_1[\tau],\ldots, s_{T-B}[\tau])$  denotes the set of \emph{non-urgent} sub-symbols for $\cC_2$~\cite{Martinian_Trott}.  
\item[Step 3] The decoder then retrieves symbols $\bs^U[\tau] = (s_{T-B+1}[\tau],\ldots, s_{T}[\tau])$ for $i-\al B \le \tau < i$ at time $t = \tau + T^\star_2$ using the parity check symbols $\bp^B[t]$ and the previously decoded non-urgent symbols. The sub-symbols $\bs^U[\cdot]$ are the urgent sub-symbols of $\cC_2$~\cite{Martinian_Trott}.
\end{enumerate}
Steps 1 and 3 above follow  from the single user SCo construction of $\cC_1$ and $\cC_2$ respectively.  Step 2 is established~\cite{FullPaper} via a recursive decoder stated in Lemma~\ref{lem:iter}. In the following we   define $\bd^A_i \!=\!(s_1[i],\ldots, s_T[i+T-1])$, $\bd^B_i \!= \!(s_1[i],\ldots,s_T[i\! -( T\! -1)\ell\!])$  as the sub-symbols involved in~\eqref{eq:pA} and \eqref{eq:pB} respectively.  
\begin{lem}\label{lem:iter}The decoder for user 2 recovers the non-urgent symbols $\bs^N[\cdot]$ in the following order
\begin{enumerate}
\item Recover the non-urgent symbols in $\bd^B_{i-\al B},\ldots, \bd^B_{i-B-1}$ using the parity check symbols $\{\bp^B[t-\Delta]\}_{t=i+T}^{\CapTau-1}$;
\item Recover the parity checks $\bp^A[i],\ldots, \bp^A[i+T-1]$ from $\bq[i],\ldots,\bq[i+T-1]$.
\item Recover the non-urgent symbols in $\bd^A_{i-1},\ldots, \bd^A_{i-B}$ using the parity checks $\bp^A[i],\ldots, \bp^A[i+T-1]$. 
\item For each $k \in \{1,\ldots, T-B-1\}$ recursively recover the remaining non-urgent symbols as follows:
\begin{itemize}
\item[(Ind. 1)] Recover the non-urgent sub-symbols in $\bd^A_{i-B-k}$ using the non-urgent sub-symbols in $\{\bd^B_j\}_{j \le i + (k-1)(\al-1)-B-1}$ and parity checks $\bp^A[\cdot]$ between $i\le t< i+T$.
\item[(Ind. 2)] Recover the non-urgent sub-symbols in $\bd^B_{i-B+(k-1)(\al-1)},\ldots, \bd^B_{i-B+k(\al-1)-1}$ using $\{\bd^A_j\}_{j\ge i-B-(k-1)}$ and the parity checks  $\bp^B[\cdot]$ between $i+T \le t < \CapTau$. 
\end{itemize}
Once this recursion terminates, all the non-urgent sub-symbols $\{\bs^N[\tau]\}_{\tau=i-\al B}^{i-1}$ are recovered by time $\CapTau-1$.
\end{enumerate}
\end{lem}
Fig.~\ref{DecodingExample} illustrates the DE-SCo $\{(4,7), (8,18)\}$ construction. Each column represents one time-index between $[-8,17]$ shown in the top row of the table.  We assume that a burst-erasure occurs between time $[-8,-1]$ and only show the relevant symbols and parity-checks. Each source symbol is split into seven sub-symbols, each occupying one row. Each source sub-symbol has two labels - one number and one letter. The letter represents the main diagonal that passes through the sub-symbol e.g., sub-symbols in $\bd^A_{-1}$ are marked $a$. The number represents the off-diagonal that passes through the sub-symbols e.g., sub-symbols in $\bd^B_{-1}$ are marked $8$. The next four rows denote the parity check sub-symbols.  The parity checks for $\cC_1$, generated by diagonal $\bd^A_i$, (c.f.~\eqref{eq:pA}) are marked by the same letter.  The four top rows, in lighter font, show the parity checks generated by the diagonal $\bd^B_i$ for $\cC_2$,  shifted by $T+B=11$ slots. These parity checks are combined with the corresponding parity checks of $\cC_1$ as shown in Fig.~\ref{DecodingExample}. For example the cell marked $(c1)$ at time $t=4$ indicates that this sub-symbol of $\bq[4]$ results by adding the parity check of $\cC_1$, marked $c$, with the parity check of $\cC_2$, marked $1$, at $t=4$. 

We illustrate the decoding steps in Lemma~\ref{lem:iter}. By construction of $\cC_1$ all the parity checks $\bp^A[t]$ after time $t\ge 7$ do not involve the erased symbols. In particular the parity checks marked by $a$, $b$ and $c$ at $t \ge 7$  can be canceled to recover  parity checks marked by $1$, $2$, $3$ and more generally all $\bp^B[t]$ for $t\ge 7$. Note that this allows us to recover the non-urgent (i.e., the top three rows) erased sub-symbols in $\bd^B_{-8},\ldots, \bd^B_{-5}$  by $t=9$. Next, compute parity checks marked by $1$, $2$, $3$ of $\cC_2$ in the shaded area in Fig.~\ref{DecodingExample} and cancel them and recover the parity checks marked by $a$, $b$ and $c$. At this point we have  all the parity checks $\bp^A[\cdot]$ for $0\le t\le 6$. Since the diagonals $\bd^A_{-1},\ldots, \bd^A_{-4}$ involve four or fewer erasures we can now recover these sub-symbols.

The remaining non-urgent sub-symbols need to be recovered in a recursive manner. Note that $\bd_{-5}^A$, marked by $e$, has five erased symbols. However the first symbol marked by $(4e)$ also belongs to $\bd^B_{-5}$  and has already been recovered. The remaining four sub-symbols can be recovered by the four available parity checks of $p_A[\cdot]$ marked by $e$. Similarly  $\bd^B_{-4}$, marked by $5$, also has five erasures, but the first symbol $(5d)$ also belongs to $\bd^A_{-4}$ and has been recovered. Hence the remaining parity checks can be recovered using the parity checks  of $\bp^B[\cdot]$. Of these, by construction of $\cC_2$, the non-urgent symbols will be recovered by time $\CapTau = 9$.  The decoder then recovers $\bd^A_{-6}$ and $\bd^B_{-3}$ in the next step to recover all non-urgent sub-symbols.

\section{Optimal MU-SCo constructions }
The DE-SCo construction in the previous section simultaneously satisfies two receivers with parameters $(B_1,T_1)$ and $B_2 = \al B_1$, and $T_2^\star = \al T_1 + B_1$ for any integer $\al > 1$ with a rate $R = \frac{T_1}{T_1+B_1}$. From the single user capacity bound, clearly this is the maximum possible rate for any MU-SCo code. Furthermore $DE-SCo$ is also optimal for any other receiver with $T_2 > T_2^\star$. Below we state the regimes in which the optimal MU-SCo constructions are known~\cite{FullPaper}. \begin{thm}
Suppose that the burst-delay parameters for user $1$ and $2$ are $(B_1, T_1)$ and $(B_2,T_2)$ respectively. Suppose that $B_2 = \al B_1$ for some integer $\al > 1$. Then optimal MU-SCo constructions are known for the following values of $T_2$:
\begin{equation}
C= \begin{cases}
\label{eq:Cap}
\frac{T_1}{T_1+B_1}, & T_2 \ge  T_2^\star \stackrel{\Delta}{=}\al T_1 + B_1,\\
\frac{T_2-B_1}{T_2-B_1+B_2}, & \max\{B_2,T_1\}+B_1\le  T_2\le T_2^\star,\\
\frac{T_1}{T_1+B_2},&T_1 \le T_2 \le T_1 + B_1, B_2 \le T_1\\
\frac{T_2}{T_2+B_2},&T_2 \le T_1
\end{cases}
\end{equation}  
\end{thm}
\begin{figure}
		\centering
		\resizebox{\columnwidth}{!}{\includegraphics[trim = 0mm 10mm 0mm 90mm, clip]{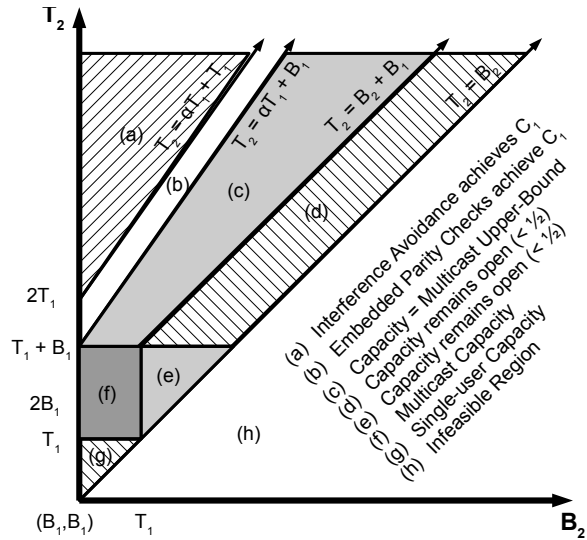}}
		\caption{Achievable Capacity Regions. Suppose $B_1$ and $T_1$ are constants, so the regions depend on the relation between $T_2$ and $B_2$.}
		\label{Regions}
\end{figure}
The different regions in~\eqref{eq:Cap} are also summarized in Fig.~\ref{Regions}. For clarity we ignore the integer proportionality constraints on the parameters in this figure (extension for non-integer values is included in \cite{FullPaper}). The first case in~\eqref{eq:Cap} corresponds to regions marked (a) and (b). In region (a), the capacity can be achieved by a simpler interference avoidance construction (illustrated in Table.~\ref{Code1225}(a)) rather than the DES-Co construction. Region (c) corresponds to case $2$ in~\eqref{eq:Cap}. The capacity is achieved through a DES-Co construction for a modified value of $T_1$. The example of $\{(1,2)-(2,4)\}$  in Section~\ref{sec:Example} falls in this region. The full derivation is provided in~\cite{FullPaper}. Region (f) corresponds to case $3$ in~\eqref{eq:Cap}. The capacity is achieved by a $(B_2,T_1)$ SCo code and finally for region (g), which corresponds to the last case in~\eqref{eq:Cap}, the capacity is achieved by a $(B_2,T_2)$ SCo code. In regions (d) and (e) the capacity remains open except for a special case when $T_2 = B_2$ and $T_1=B_1$ where the concatenation based scheme is known to be optimal~\cite{Khisti}.

\section{Conclusion}

This paper constructs a new class of streaming erasure codes that do not commit apriori to a given delay, but rather achieve a delay based on the channel conditions. We model this setup as a multicast problem to two receivers whose channels introduce different erasure-burst lengths and require different delays. The DE-SCo construction embeds new parity checks into the single-user code, in a such a way that the weaker receiver can also recover the stream with an information theoretically optimum delay.  MU-SCo constructions that are optimal for a range of burst-delay parameters are also provided.

At a broader level, our paper sheds insight into the interesting impact delay can have in multiuser channels. How can one order two receivers --- one with a weaker channel but a relaxed delay constraint and one with a stronger channel but a stringent delay requirement?  The present paper  illustrates the intricate interaction between delay  and channel quality. We hope that this paper sparks further interest into this fascinating area.

\end{document}